\documentclass[doublecol]{epl2}
\pdfoutput=1 
\usepackage[charter]{mathdesign}
\usepackage[amssymb,mediumspace,mediumqspace]{SIunits}
\usepackage[font=footnotesize,labelfont=bf,singlelinecheck=false]{caption}
\usepackage{sidecap}
\usepackage[colorlinks,plainpages=false,linkcolor=blue,urlcolor=blue,citecolor=blue,pdfpagemode=UseNone,pdfstartview=FitBH]{hyperref}

\newcommand{\vecta}[1]{\mathbf{#1}}
\newcommand{\half}{\frac{1}{\protect\raisebox{0.8pt}{\scriptsize 2}}}
\newcommand{\quarter}{\frac{1}{\protect\raisebox{0.8pt}{\scriptsize 4}}}

\newcommand{\bfca}{\mbox{BaFe$_{2-x}$Co$_x$As$_2$\,}}

\newcommand{\afsx}{\mbox{\textit{A}$_x$Fe$_{2-y}$Se$_2$}\,}
\newcommand{\afsi}{\mbox{\textit{A}$_{0.8}$Fe$_{1.6}$Se$_2$}\,}
\newcommand{\afsii}{\mbox{\textit{A}$_x$Fe$_2$Se$_2$}\,}
\newcommand{\afsiii}{\mbox{\textit{A}Fe$_2$Se$_2$}\,}
\newcommand{\rfsx }{\mbox{Rb$_{x}$Fe$_{2-y}$Se$_2$}\,}

\newcommand{\rfsi}{\mbox{Rb$_{0.8}$Fe$_{1.6}$Se$_2$}\,}

\newcommand{\kfsx}{\mbox{K$_x$Fe$_{2-y}$Se$_2$}\,}
\newcommand{\cfsx}{\mbox{Cs$_x$Fe$_{2-y}$Se$_2$}\,}
\newcommand{\kfsii}{\mbox{K$_x$Fe$_{2}$Se$_2$}\,}
\newcommand{\kfsi}{\mbox{K$_{0.8}$Fe$_{1.6}$Se$_2$}\,}

\title{Conformity of spin fluctuations in alkali-metal iron selenide superconductors inferred from the observation of a magnetic resonant mode in \kfsx}
\shorttitle{The magnetic resonant mode in \kfsx} 

\author{G.\,Friemel\inst{1} \and W.\,P. Liu\inst{1} \and E.\,A.\,Goremychkin\inst{2} \and Y.\,Liu \inst{1} \and J.\,T. Park \inst{1,3} \and O.\,Sobolev \inst{3} \and C.\,T. Lin \inst{1} \and B.\,Keimer \inst{1} \and D.\,S.\,Inosov\inst{1}\hspace{-5pt}\thanks{E-mail: \email{d.inosov@fkf.mpg.de}}}
\shortauthor{G. Friemel \etal}

\institute{
  \inst{1} Max Planck Institute for Solid State Research, Heisenbergstra{\ss}e~1, D-70569 Stuttgart, Germany.\\
  \inst{2} ISIS Facility, STFC Rutherford Appleton Laboratory, Oxfordshire OX11 0QX, United Kingdom.\\
  \inst{3} Forschungsneutronenquelle Heinz Maier-Leibnitz (FRM-II), TU München, D-85747 Garching, Germany.
}

\abstract{
Spin excitations stemming from the metallic phase of the ferrochalcogenide superconductor K$_{0.77}$Fe$_{1.85}$Se$_2$ ($T_\text{c}=32\,\mathrm{K}$) were mapped out in the \textit{ab} plane by means of the time-of-flight neutron spectroscopy. We observed a magnetic resonant mode at $\vecta{Q}_\text{res}=(\half \quarter)$, whose energy and in-plane shape are almost identical to those in the related compound Rb$_{0.8}$Fe$_{1.6}$Se$_2$. This lets us infer that there is a unique underlying electronic structure of the bulk superconducting phase \kfsii, which is universal for all alkali-metal iron selenide superconductors and stands in contrast to the doping-tunable phase diagrams of the related iron pnictides. Furthermore, the spectral weight of the resonance on the absolute scale, normalized to the volume fraction of the superconducting phase, is several times larger than in optimally doped \bfca. We also found no evidence for any additional low-energy branches of spin excitations away from $\mathbf{Q}_{\rm res}$. Our results provide new input for theoretical models of the spin dynamics in iron based superconductors.
}

\begin{document}
\maketitle

\section{Introduction}
The discovery of the alkali-metal iron selenide superconductors \afsx (\textit{A}=K, Rb, Cs) with a $T_\text{c}$ as high as $32\,\mathrm{K}$ introduced qualitatively new compounds in the research on iron-based high temperature superconductors, which differ from the analogous iron-arsenide compounds in a number of properties \cite{discovery}. First of all, their crystals tend to grow in a composition, which is deficient both in iron and in the alkali atoms, compared to the `parent' 122 stoichiometry.  They exhibit a phase transition above $500\,\mathrm{K}$ to a vacancy ordered phase with a $\sqrt{5} \times \sqrt{5}$ superstructure, corresponding to \afsi stoichiometry that hosts a checkerboard antiferromagnetic (AFM) state \cite{Shermadini11,order}. However, this phase is insulating and a number of experimental probes could show that the superconducting (SC) phase is spatially separated from this magnetic phase, with both being sandwiched along the $c$-axis \cite{tem,CharnukhaCvitkovic}. Scanning tunneling microscopy (STM) and nuclear magnetic resonance (NMR) experiments could identify the SC phase to be free of iron vacancies \cite{stm,nmr}, which, considering the valence of iron, would constitute a strongly electron-doped system, compared to the iron arsenide compounds. The Fermi surface, consisting of large electron pockets at the $M$ point, as reported by angle-resolved photoemission spectroscopy (ARPES), is consistent with this conclusion \cite{gap1,gap2}. However, the growth of a pure SC phase has not been possible yet, which prevented a systematic study of the phase diagram in dependence of the electron doping by tuning the content of the alkali atoms. All studies on the electronic properties of the superconducting \afsx compounds with varying nominal compositions of $A$ were done on a mixture between the AFM phase and the metallic/SC phase \cite{pd1,pd2}. It was suggested that the SC dome has basically a rectangular shape, extending only over a narrow region of Fe content or Fe valency and showing a constant $T_\text{c}$ of $32\,\mathrm{K}$ \cite{pd1,pd2}. On the one hand, the latter fact implies that superconductivity always originates from the same phase with a certain electron doping level (scenario I). On the other hand, superconductivity was found in a wide range of compositions, extending from K$_{0.64}$Fe$_{1.44}$Se$_2$ \cite{comp1} to K$_{0.77}$Fe$_{1.85}$Se$_2$ (this study), indicating an extended SC dome as a function of either K concentration, Fe content, or Se vacancies \cite{stm2} (scenario II). However, the separation of the sample into a majority AFM phase with 88\% volume fraction and a minority metallic/SC phase with 12\% volume fraction, as seen by muon-spin rotation spectroscopy ($\mu$SR) \cite{muSR}, makes it difficult to refine the chemical structure of the latter phase by X-ray diffraction \cite{x-ray}, precluding a reliable comparison with first-principles calculations.

Inelastic neutron scattering (INS) as a bulk sensitive probe is in the focus of the investigation towards resolving this issue as it enables the observation of the magnetic resonant mode at $\vecta{Q}_\text{res}=(\half\,\quarter)$ wave vector and at an energy of $\hbar\omega_\text{res}=14\,\mathrm{meV}$  in the SC state of \rfsi single crystals \cite{ParkFriemel11}. The resonant intensity is two dimensional, having an elliptical in-plane shape \cite{FriemelPark12}. This feature could be reproduced by theoretical calculations of the imaginary part of the dynamical spin susceptibility in the framework of the random phase approximation (RPA), starting from the vacancy free \afsiii \cite{Maier11,FriemelPark12}. Here, the position of the resonance corresponds to the nesting vector of the Fermi surface connecting the flat parts of the electron pockets, which is also evidenced by the sizeable magnetic normal-state response at $T=35\,\mathrm{K}$ \cite{FriemelPark12}. In order to match the experimentally determined nesting vector, we had to apply a rigid band shift, corresponding to an electron doping of 0.18 electrons/Fe, which would imply a chemical formula of Rb$_{0.36}$Fe$_2$Se$_2$ under the assumption of vacancy-free FeSe planes. A recent NMR experiment on a similar sample, with the nominal composition Rb$_{0.74}$Fe$_{1.6}$Se$_2$, could  estimate a similar value of $x=0.29$ by comparing the spectral weight of the NMR signal from Rb atoms belonging to the SC phase and the magnetic phase \cite{nmr}. The good agreement among the independent results from ARPES, NMR, and INS indicates that INS can serve as a tool to indirectly probe the size of the Fermi surface by measuring the wave vector of the resonant mode, which represents the nesting vector. This means, application of a larger or smaller doping level to the SC phase should expand or shrink the electron pockets, respectively, shifting the nesting vector to $\vecta{Q}=(\half,\,\quarter\pm\delta)$, where `+' corresponds to reduced and `$-$' to increased doping in comparison to Rb$_{0.36}$Fe$_2$Se$_2$.

Another recent INS study on Rb$_{0.82}$Fe$_{1.68}$Se$_2$ by time-of-flight (TOF) spectroscopy not only confirmed the resonant mode at exactly the same wave vector, but also reported even stronger branches of incommensurate spin excitations up to $\hbar\omega=16\,\mathrm{meV}$ around $\vecta{Q}=(\piup,0)$, which exhibit no change of intensity upon entering the SC state \cite{WangLi12}. Due to the absence of hole pockets at the $\Gamma$ point and the consequent suppression of the $(\piup,0)$ scattering channel for particle-hole excitations, these branches were interpreted as originating from localized spins, which, together with itinerant excitations, were claimed to be important ingredients for high-$T_\text{c}$ superconductivity \cite{WangLi12}.

Here, we present a TOF study of the normal-state excitations and of the magnetic resonant mode in a superconducting \kfsx compound, with a Fe content of $1.81<2-y<1.89$ and a K content of $0.74<x<0.8$, as determined by energy-dispersive X-ray spectroscopy (EDX). Despite the significant deviation in iron content from the previously studied \rfsi sample, we observe the normal-state intensity and the resonant mode at almost exactly the same position in the Brillouin zone. This indicates that the nesting condition and consequently the Fermi surface are identical to those in the \rfsi sample and provides strong evidence in favor of the electronically unique SC phase with a pinned doping level for all superconducting \afsx compounds, independently of the type of alkali atom $A$ or the experimentally determined average stoichiometry of $A$ and Fe.

\begin{figure}[t]
\includegraphics[width=0.9\columnwidth]{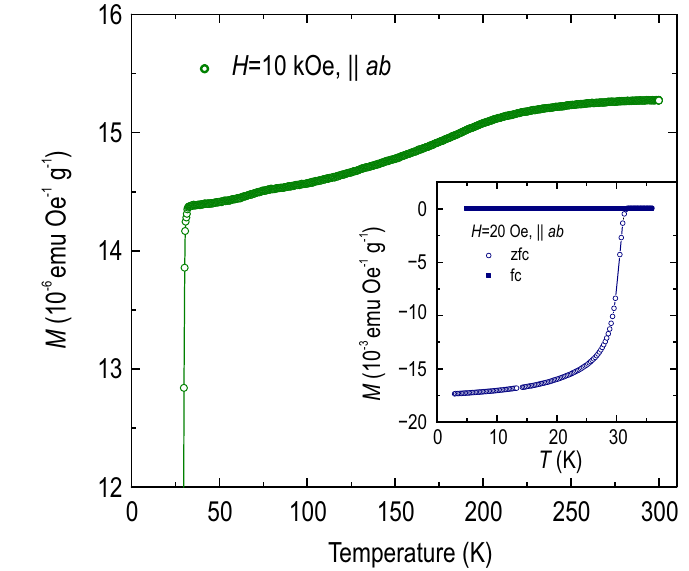}

\vspace{-0.5em}
\caption{Temperature dependence of the magnetization measured in a magnetic field of $H=10\,\mathrm{kOe}$, parallel to the basal plane (\textit{ab}-plane) of the sample. Inset: Field-cooled and zero-field-cooled magnetization curves measured in a magnetic field of $H=20\,\mathrm{Oe}$.}
\label{fig:magn}

\vspace{-1.5em}
\end{figure}

\begin{figure*}[t]

\vspace{-1em}
\begin{minipage}[t]{0.7\textwidth}

\vspace{0pt}
\includegraphics[width=\textwidth]{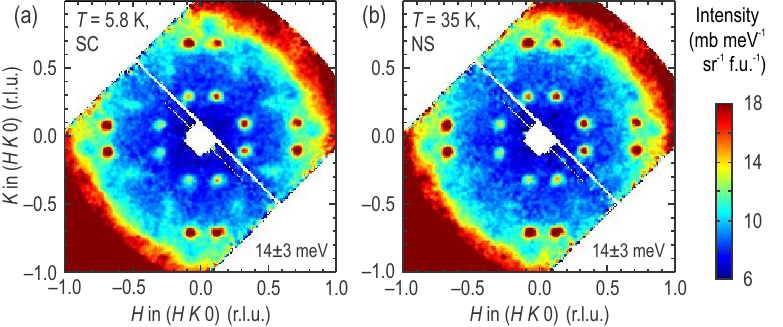}
\vspace{-0.5em}
\end{minipage} \hfill
\begin{minipage}[t]{0.27\textwidth}

\vspace{0pt}
\caption{In-plane wave-vector dependence of the spin excitations in \kfsx, integrated over the energy range of $\hbar\omega=14\pm 3\,\mathrm{meV}$, obtained with an incident neutron energy of $E_\text{i}=35\,\mathrm{meV}$ in the (a) SC state at $T=5.8\,\mathrm{K}$ and (b) normal state at $T=35\,\mathrm{K}$. The out-of-plane momentum at $\vecta{Q}_\text{res}=(0.5,0.25)$ is $L=1.33\,\mathrm{r.l.u.}$}
\label{fig:tofraw}
\end{minipage}

\vspace{-1.5em}
\end{figure*}

\section{Experimental details}
The sample used in this study consists of a mosaic of \kfsx single crystals with a mass of $2.7$\,g, which were grown by the floating-zone technique and coaligned to a mosaicity of $2.8\,\degree$. The details of the growth procedure and characterization data can be found in Ref.\,\citenum{growth}. All samples were characterized by magnetization measurements in a vibrating-sample SQUID magnetometer. As displayed in Fig.\,\ref{fig:magn} and its inset, the samples show appreciable screening of the magnetic field below $T_\text{c}=32\,\mathrm{K}$ for zero-field-cooling, as well as a monotonic normal-state susceptibility vs. temperature up to $300\,\mathrm{K}$, consistent with literature \cite{pd3}.
INS experiments were carried out at the MERLIN time-of-flight chopper spectrometer at the Rutherford-Appleton Laboratory, Didcot, UK. The incident neutron beam had an energy of $E_\text{i}=35\,\mathrm{meV}$ and was directed along the $\mathbf{c}$-axis. The intensity was normalized to units of the differential cross section $\text{d}^2\sigma/\text{d}\Omega\text{d}\omega$ ($\mathrm{mb\, meV^{-1}sr^{-1}f.u.^{-1}}$) by means of a vanadium standard. Here f.u. stands for the formula unit of K$_{0.77}$Fe$_{1.85}$Se$_2$, which is the average composition of the sample.  Triple-axis spectroscopy (TAS) measurements on the same sample were performed at the PUMA spectrometer (FRM-II, Garching) with a fixed final neutron wave vector of $k_\text{f}=\unit{2.662}{\reciprocal\angstrom}$. We used double-focused pyrolytic graphite (PG) (002) for both the monochromator and the analyzer and installed two PG filters between the sample and analyzer to suppress contamination from higher harmonics of the neutron wave length. The presentation and analysis of the inelastic scattering intensity refers to the Brillouin zone of the iron sublattice, which has the lattice parameters $a=b=\unit{2.75}{\angstrom}$ and $c=\unit{7.0}{\angstrom}$. This notation is consistent with previous INS reports \cite{ParkFriemel11,FriemelPark12}.

\section{Experimental results}
\begin{figure}[t]

\vspace{-1em}
\centering
\includegraphics[width=0.75\columnwidth]{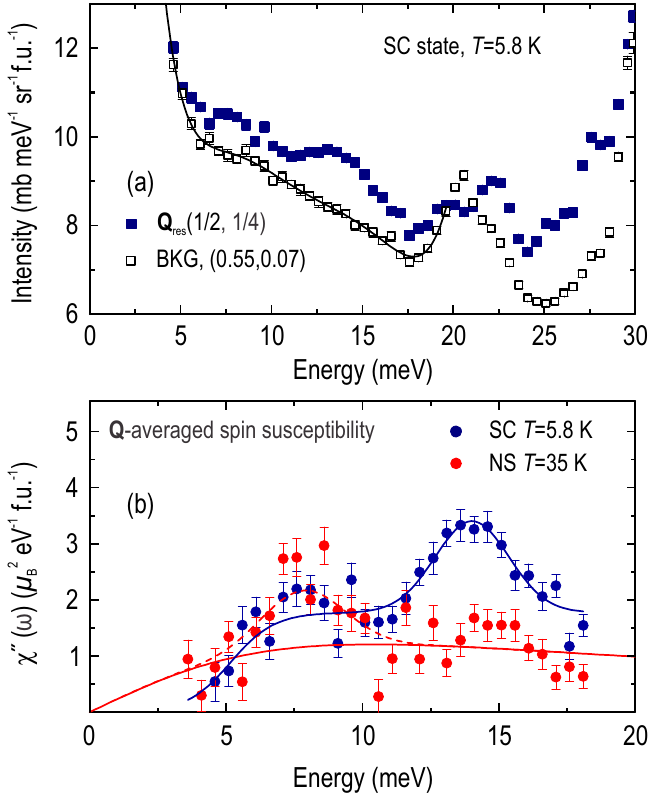}
\caption{(a) Spectrum in the SC state at the wave vector of the resonant mode, $\vecta{Q}_\text{res}=(\half\,\quarter)$. The intensity was integrated over $H=0.5\pm0.1$ and $K=0.25\pm0.05$. The background intensity vs. energy was taken at $(0.55\pm0.06,0.07\pm0.07)$. (b) $\vecta{Q}$-averaged dynamical spin susceptibility, $\chi\sp{\prime\prime}(\omega)$, in the SC ($T=5.8\,\mathrm{K}$) and normal ($T=35\,\mathrm{K}$) states.}
\label{fig:spec}

\vspace{-1.5em}
\end{figure}

In Figs.\,\ref{fig:tofraw} (a) and \ref{fig:tofraw} (b) we present the in-plane wave-vector dependence of the TOF scattering intensity, projected on the $(HK0)$ plane and integrated over the energy range of $\hbar\omega=14\pm3\,\mathrm{meV}$ for the SC and normal states, respectively. The strong excitations emerging from $(0.3,\pm0.1)$, $(0.7,\pm0.1)$ and equivalent wave vectors are spin wave excitations centered at the magnetic Bragg peak positions of the insulating/\,antiferromagnetic phase, which have been recently studied in detail \cite{spinwave}. Here we are interested in the weaker feature centered at $\vecta{Q}_\text{res}=(\half\,\quarter)$ in Fig.\,\ref{fig:tofraw} (a) and equivalent positions, which takes the shape of an ellipse. A significant decrease of the intensity in the normal state strongly suggests that this excitation is the resonant mode which was previously found in the related \rfsi compound \cite{ParkFriemel11}. To investigate this finding in more detail, in Fig.\,\ref{fig:spec}(a) we show an energy cut, integrated over the region of the ellipse at $H=(0.5\pm0.1)\,\mathrm{r.l.u.}$ and $K=(0.25\pm0.05)\,\mathrm{r.l.u.}$ An estimate of the background was taken near $\vecta{Q}_\text{bkg}=(0.55,0.07)$, which has the same absolute $|\vecta{Q}|$ as the wave vector of the resonance. Fig.\,\ref{fig:spec}(b) shows the $\vecta{Q}$-averaged imaginary part of the spin susceptibility $\chi\sp{\prime\prime}(\omega)=\int \chi\sp{\prime\prime}(\vecta{Q},\omega)\,\text{d}\vecta{Q}\big/ \int \text{d}\vecta{Q}$, where $\chi\sp{\prime\prime}(\vecta{Q},\omega)$ can be obtained from the scattering function $S(\vecta{Q},\omega)$ after correcting for the Bose factor and the Fe$^{2+}$ magnetic form factor. Here the averaging is performed by normalizing the total spectral weight of four resonant peaks at symmetrically equivalent positions in the unfolded Brillouin zone by the total Brillouin zone area. A peak is visible around $\hbar\omega_\text{res}=14\,\mathrm{meV}$, which coincides with the energy found for the resonant mode in \rfsi \cite{ParkFriemel11}. This excitation forms on top of a broad preexisting normal-state response measured at $T=35\,\mathrm{K}$. Due to the proper assessment of background here, the spectrum in Fig.\,\ref{fig:spec}(b) shows the total magnetic intensity on the absolute scale unlike Refs. \citenum{ParkFriemel11}, \citenum{FriemelPark12}, and \citenum{WangLi12}, where only the resonant intensity was obtained by subtracting the normal-state spectrum from the SC spectrum.

Due to the opening of the SC gap, the low energy spectral weight is depleted and redistributed to higher energies, creating the characteristic resonance peak, as predicted by an RPA based theory \cite{Maier11} and  observed in numerous related iron-based superconductors \cite{res-mode,bfca}. However, the SC spectrum in Fig.\,\ref{fig:spec}(b) does not show the full suppression to zero intensity below the resonance energy, but rather a shoulder or a plateau, which reaches down to $5\,\mathrm{meV}$. A spin gap, showing zero intensity, can be assumed for energies smaller than $4\,\mathrm{meV}$. However, these energies are not accessible here due to the strong contamination from  incoherent scattering centered at the elastic line. Especially the normal-state spectrum suggests that there is an additional low energy (LE) excitation around $7.5\,\mathrm{meV}$, which can be reasonably fitted with a damped harmonic oscillator function plus a Gaussian function to account for the peak. As we will show in the Appendix, this additional feature overlaying the spin-gap region is most probably of nonmagnetic origin and is not part of the spin-excitation spectrum.

\begin{figure}[t]

\vspace{-1em}
\includegraphics[width=\columnwidth]{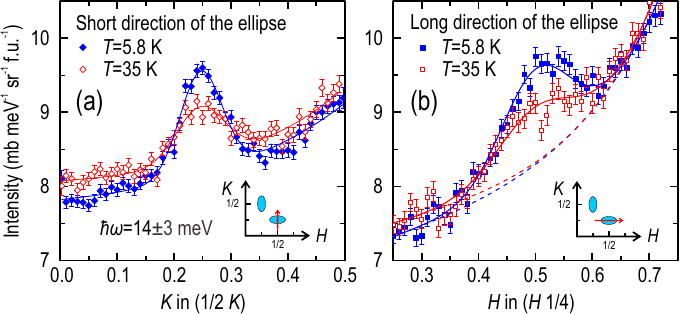}
\caption{(a) Momentum cuts through the resonance peak and the corresponding normal-state profile at $\hbar\omega=14\pm3\,\mathrm{meV}$ along  the short direction of its cross section, integrated over $0.5\pm0.1$ in the $H$ direction. (b) Similar momentum cuts along the long direction of the ellipse integrated over $0.25\pm0.05$ in the $K$ direction.}
\label{fig:R-cuts}

\vspace{-1.5em}
\end{figure}

In Figs.\,\ref{fig:R-cuts} (a) and \ref{fig:R-cuts}(b) we plot momentum cuts through the short and long directions of the ellipse, respectively, which are integrated over the energy window $\hbar\omega=14\pm3\,\mathrm{meV}$. The peak widths in both directions give an aspect ratio of $1.63$, confirming the anisotropic in-plane cross section of the spin fluctuations \cite{FriemelPark12}. The $K$ coordinate of the wave vector of the resonant mode is particularly interesting, as it can be related to the nesting vector of the Fermi surface. From a Gaussian fit in Fig.\,\ref{fig:R-cuts}(a) we obtain $K=(0.247\pm0.002)\,\mathrm{r.l.u.}$, i.e. almost perfectly commensurate within the experimental uncertainty, and a full width of half maximum (FWHM) of $0.089\,\mathrm{r.l.u.}$ in the SC state. From an analogous momentum scan on the triple-axis spectrometer PUMA in the $(HK0)$ scattering plane (not shown) we derived similar values: $K=(0.241\pm0.005)\,\mathrm{r.l.u.}$ and $\text{FWHM}=0.082\,\mathrm{r.l.u.}$ These values perfectly agree with those of \rfsi with $K=(0.244\pm0.002)\,\mathrm{r.l.u.}$ at $L=0.5$ and a FWHM of $0.078\,\mathrm{r.l.u.}$  \cite{FriemelPark12}. The good overall agreement of the peak widths independent of the instrument indicates that the broadening of the resonance peak is mainly intrinsic, and that the instrumental resolution has at most a minor effect. This is also supported by the comparable widths of the sharper peak (not shown), which one obtains choosing the energy window as $\hbar\omega=12\pm1\,\mathrm{meV}$:  $(0.047\pm0.009)\,\mathrm{r.l.u.}$ here and $(0.052\pm0.009)\,\mathrm{r.l.u.}$ in Ref.\,\citenum{FriemelPark12}. 

Given that the wave vector of the resonance coincides with the nesting vector of the Fermi surface, we would expect an additional broadening due to the wider spread in composition for the \kfsx sample, as compared to the sample in Ref.\,\citenum{FriemelPark12}, resulting in an inhomogeneous doping level. However, the absence of such an effect lets us conclude that the doping level is uniform for the SC phase and that differences in compositions for the different crystals might be related to different fractions of the SC phase, having \kfsii composition \cite{nmr}, and the antiferromagnetic/\,insulating phase with \kfsi composition.

Having the inelastic intensity converted to absolute units, it is worthwhile to estimate the spectral weight of the resonance peak. The $\vecta{Q}$-averaged spin susceptibility  in Fig.\,\ref{fig:spec}\,(b) corresponds to the local susceptibility, which is a measure of the fluctuating moment. By integrating the difference of the SC- and normal-state spectra, one obtains a resonant spectral weight of $\int  (\chi_\text{SC}\sp{\prime\prime}-\chi_\text{NS}\sp{\prime\prime})\, \text{d}\omega =(0.011\pm0.003)\,\mu_\text{B}^2/ \mathrm{f.u.}$ This result is comparable with the value of the resonant intensity in optimally doped \bfca (BFCA) \cite{bfca}.
However, taking into account the SC volume fraction of only 12--20\% \cite{muSR,CharnukhaCvitkovic}, our estimate implies a total resonantly enhanced spectral weight in the SC phase of \kfsx that is at least 2--3 times larger than in iron pnictides, which are known to be bulk superconductors.

Another recent TOF study on a superconducting \rfsx sample reported a set of incommensurate excitations at $\hbar\omega=8\,\mathrm{meV}$  around $\vecta{Q}=(\piup,0)$, which are 16 times more intense than the magnetic resonant mode \cite{WangLi12}. These excitations merge and become commensurate at $\hbar\omega=16\,\mathrm{meV}$. They exhibit no change of intensity between the SC and the normal states and persist at temperatures up to $250\,\mathrm{K}$. The dispersion of these excitations would cross the energy window $\hbar\omega=14\pm3\,\mathrm{meV}$ of the  $(H,K)$-plane cut presented in Fig.\,\ref{fig:tofraw}. However, we clearly do not observe these excitations. We argue that they might emerge from an unknown third phase with no relationship to superconductivity. A candidate is the recently discovered semiconducting phase in \kfsx, which exhibits $(\piup,0)$-stripe AFM order below $T_\text{N}=280\,\mathrm{K}$ \cite{ZhaoCao12} and comprises a rather large magnetic moment of $2.8\,\mu_\text{B}$. An additional dynamical modulation with a small wave number could explain the incommensurability of the $(\piup,0)$ excitations at lower energies around $8\,\mathrm{meV}$ \cite{WangLi12}.

\section{Discussion}
Our inelastic neutron scattering data show convincingly the development of a spin resonance peak in the SC state of \kfsx, which coincides in momentum position and energy with the spin resonance peak in \rfsx \cite{ParkFriemel11,FriemelPark12}. Its spectral weight, when normalized to the volume fraction of the SC phase, considerably exceeds that in optimally doped BFCA. In order to explain this observation in the nesting scenario, an investigation of the spectral weight transfer below $T_\text{c}$ from low energies to energies around $\hbar\omega_\text{res}$ is necessary. The imaginary part of the normal-state susceptibility $\chi_\text{NS}\sp{\prime\prime}$ depends linearly on energy transfer, $\hbar\omega$, for small $\omega$. This spectral weight is redistributed upon opening the SC gap, thereby obeying the sum rule for the energy- and $\textbf{Q}$-integrated scattering intensity $S=\int_{-\infty}^{\infty}\mathrm{d}\omega\int \mathrm{d}\vecta{Q}\, S(\vecta{Q},\omega)$. Given the relation $\chi_\text{NS}\sp{\prime\prime}=~(1-e^{-\hbar\omega/k_\text{B}T})S(\vecta{Q},\omega)$, a larger resonance peak implies that the normal-state  intensity at small $\omega$ in \kfsx must be fairly larger than in BFCA, $ (\chi_\text{NS}\sp{\prime\prime}/\omega)_\text{KFS}>(\chi_\text{NS}\sp{\prime\prime}/\omega)_\text{BFCA}$. Even a larger spin gap in \kfsx, suggested by its $\sim$\,40\% higher resonance energy, as compared to BFCA, can not account for such a substantial difference. The term $\chi_\text{NS}\sp{\prime\prime}/\omega$ at small energies can be theoretically estimated from RPA-based calculations of the renormalized Lindhard function taking into account the real band structure, which would allow another test for the applicability of the nesting scenario \cite{Maier11,ParkInosov10}.

This is important with respect to the fact that the wave vector of the magnetic intensity seems to be centered at the commensurate position $\vecta{Q}_\text{res}=(\half\,\quarter)$, independent of  the alkali element or the actual composition of the sample. This result was recently confirmed also for the \cfsx compound \cite{cfs}. In a rigid-band picture one would expect a shift of the nesting vector upon shifting the chemical potential and thus contracting or expanding the electron pockets. Instead, we observe that the SC phase appears to be pinned around a certain doping level of 0.15 electrons per Fe ion \cite{FriemelPark12,nmr}, which is anomalous knowing that iron pnictides and selenides usually form extended SC domes in dependence on charge or isovalent doping. On the other hand, such a pinning of the doping level might also be a consequence of the chemical structure of the SC phase \afsii, which has fully occupied Fe sites. This phase is embedded in a matrix of the majority $\sqrt{5} \times \sqrt{5}$  antiferromagnetic/insulating phase \cite{SpellerBritton12,CharnukhaCvitkovic}. To our knowledge there exists no chemically pure SC phase without the coexisting $\sqrt{5} \times \sqrt{5}$ phase in bulk single crystals, which may indicate that the latter is necessary to stabilize the SC/metallic phase. NMR experiments estimated a Rb content of only $x=0.29$, whereas a refinement of neutron powder diffraction data provided an estimate of $x=0.6$ \cite{PomjakushinPomjakushina12}. Both studies could not distinguish whether the Rb atoms exhibit correlations in the \textit{ab} plane \cite{x-ray} or if they are randomly distributed. Further microstructural studies are needed to elucidate this question, which is crucial to understand the electronic structure and the origin of the high-$T_\text{c}$ superconductivity in the \afsx compounds.

\vspace{-1em}
\acknowledgments
We acknowledge discussions with A. Boothroyd, S. V. Borisenko, A. V. Chubukov, P. Dai, I. I. Mazin, A. E. Taylor and A. N. Yaresko. This work has been supported, in part, by the DFG within the SPP 1458, under Grant No. BO3537/1-1.

\begin{figure}[t]
\includegraphics[width=\columnwidth]{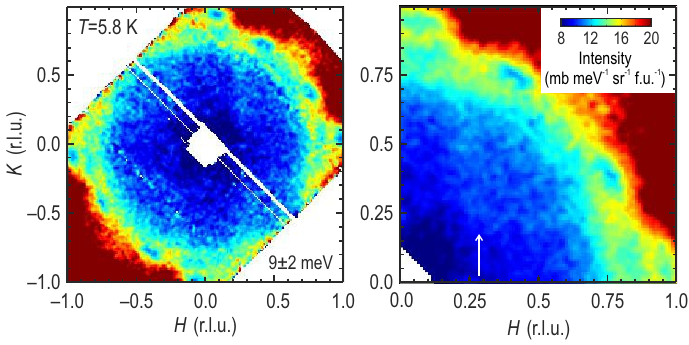}
\caption{(a) In-plane wave-vector dependence of the spin excitations in \kfsx, integrated over the energy range of $E=9\pm 2\,\mathrm{meV}$ in the SC state at $T=5.8\,\mathrm{K}$.  (a) Raw map of the accessible part of the Brillouin zone.  (b) The same map symmetrized with respect to the symmetry axes $(100)$ and $(010)$.}
\label{fig:LEmap}

\vspace{-1em}
\end{figure}

\begin{figure}[t]
\includegraphics[width=\columnwidth]{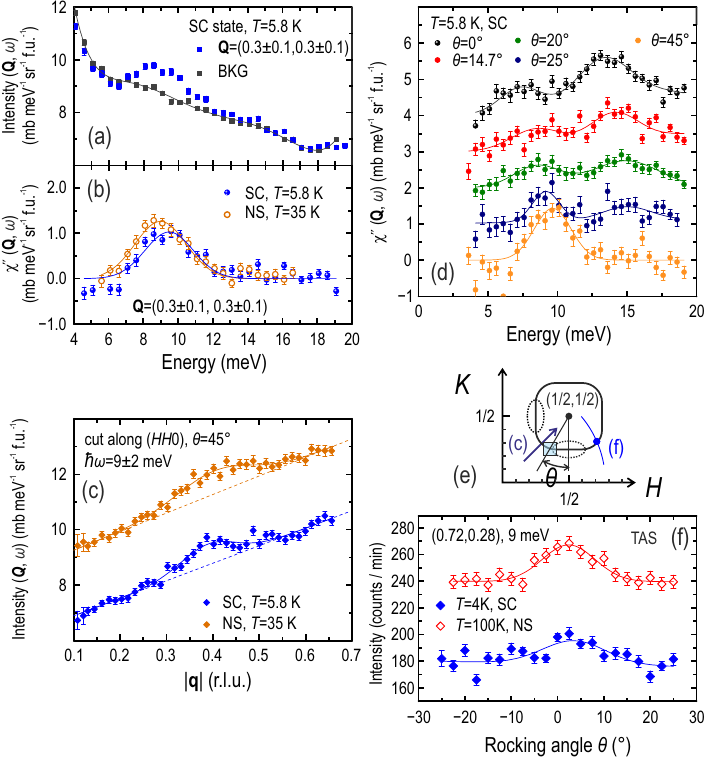}
\caption{(a) Energy cuts integrated in the region between the resonance positions at $\vecta{Q}=(0.3\pm0.1,0.3\pm0.1)$ and background intensity vs. energy. (b) Background and Bose-factor corrected intensity of the energy cut in (a) for the SC and normal states. (c) Momentum cuts along $(H\,H\,0)$ at the LE mode energy $\hbar\omega=9\pm2\,\mathrm{meV}$. (d)~Several energy cuts along the ring-shaped reciprocal space structure of the LE mode, integrated over a square with a width of $0.1\,\mathrm{r.l.u.}$ and shifted by 1 unit vertically from each other for clarity. (e) Sketch pointing out the position of the energy cuts in panel (d) and the momentum cuts in panels (c) and (f). (f) Rocking scan at $E=9\,\mathrm{meV}$ intersecting the LE mode in the second Brillouin zone.}
\label{fig:LE}

\vspace{-1.5em}
\end{figure}

\section{\textsc{Appendix}}
Beside the resonant mode,  we observed another excitation in the spectrum at $\vecta{Q}_\text{res}=(\half\,\quarter)$ in Fig.\,\ref{fig:spec}, which is visible as a shallow feature in both SC and normal states for energies between $5$ and $11 \,\mathrm{meV}$. To explore its reciprocal space structure, in Fig.\,\ref{fig:LEmap}(a) we present a $(H\,K)$  plane cut, integrated in an energy window of $9\pm2\,\mathrm{meV}$. There, one can see clouds of intensity surrounding $\vecta{Q}=(\half\,\half)$ and equivalent positions. A closer look at this in panel (b) reveals  that this intensity takes the shape of a ring, centered at $\vecta{Q}=(\half\,\half)$. It seems that its contour is also passing through the wave vector of the resonance. To highlight this more, we show an energy cut integrated over $\vecta{Q}=(0.3\pm0.1,0.3\pm0.1)$, which approximately lies on the ring contour between the equivalent resonance positions [marked by a white arrow in Fig.\,\ref{fig:LEmap}(b)] in panels (a) and (b) of Fig.\,\ref{fig:LE}. In addition, we plot a momentum cut along the $(HH0)$ direction in Fig.\,\ref{fig:LE}(c), which would intersect the ring radially. In all plots we see an excitation centered at $\hbar\omega=9\,\mathrm{meV}$ and $\vecta{Q}_\text{L}=(0.28,\,0.28)$, with an out-of-plane momentum of $L=0.75\,\mathrm{r.l.u.}$ Panel (d) in Fig.\,\ref{fig:LE} shows spectra along the contour of the ring integrated in a reciprocal space region with a width of $0.1\,\mathrm{r.l.u.}$ Starting from the diagonal position between the resonances at $\theta=45\,\degree$, the LE excitation is visible in all spectra and therefore is connected with the shallow peak around $\hbar\omega=7.5\,\mathrm{meV}$ in the spectrum at the resonance wave vector ($\theta=0\,\degree$), as seen in Fig.\,\ref{fig:spec}\,(b) and Fig.\,\ref{fig:LE}\,(d). This LE mode does not seem to show any visible response to superconductivity, as revealed by the nearly identical spectra at $\vecta{Q}_\text{L}=(0.3,\,0.3)$ in the SC and normal states in Fig.\,\ref{fig:LE}\,(b).

To investigate, whether the feature has a magnetic origin, we performed a rocking scan with a triple-axis spectrometer through the equivalent wave vector in the second Brillouin zone at $\vecta{Q}=(0.72,0.28)$ at a higher temperature.  The scan was done in the $(HK0)$ scattering plane at $\hbar\omega=9\,\mathrm{meV}$, as sketched by the blue trajectory in Fig.\,\ref{fig:LE}\,(e) and is shown in Fig.\,\ref{fig:LE}\,(f). The observed peak has an intensity comparable to the resonance peak intensity, confirming the observation of the TOF experiment. Upon warming to $T=100\,\mathrm{K}$ its intensity increases by a factor of $1.6\pm0.5$. This matches the scaling expected from the Bose factor $1/(1-e^{-\hbar\omega/k_\text{B}T})$, taking the value of 1.54 at $T=100\,\mathrm{K}$ and $\hbar\omega=9\,\mathrm{meV}$, which strongly suggests that this LE peak is due to a phonon, since one would expect a decrease of intensity if it were a magnetic excitation such as a paramagnon or a spin wave. It possibly emerges from the structural Bragg peak at $\vecta{Q}_\text{1Fe}=(\half\,\half\,\half)$, which corresponds  to $\vecta{Q}_\text{2Fe}=(1\,0\,1)$ in the conventional $I4/mmm$ notation with two iron atoms per unit cell~\cite{ParkInosov10}.

\end{document}